# Effect of eHMI on pedestrian road crossing behavior in shared space with Automated Vehicles – A Virtual Reality study [*]

Yan Feng, Haneen Farah, Bart van Arem

*Abstract*— A shared space area is a low-speed urban area in which pedestrians, cyclists, and vehicles share the road, often relying on informal interaction rules and greatly expanding freedom of movement for pedestrians and cyclists. While shared space has the potential to improve pedestrian priority in urban areas, it presents unique challenges for pedestrian-AV interaction due to the absence of a clear right of way. The current study applied Virtual Reality (VR) experiments to investigate pedestrian-AV interaction in a shared space, with a particular focus on the impact of external human-machine interfaces (eHMIs) on pedestrian crossing behavior. Fifty-three participants took part in the VR experiment and three eHMI conditions were investigated: no eHMI, eHMI with a pedestrian sign on the windshield, and eHMI with a projected zebra crossing on the road. Data collected via VR and questionnaires were used for objective and subjective measures to understand pedestrian-AV interaction. The study revealed that the presence of eHMI had an impact on participants' gazing behavior but not on their crossing decisions. Additionally, participants had a positive user experience with the current VR setting and expressed a high level of trust and perceived safety during their interaction with the AV. These findings highlight the potential of utilizing VR to explore and understand pedestrian-AV interactions.

## I. INTRODUCTION

The development of automated vehicles (AVs) offers the potential to reduce traffic congestion, improve traffic safety, and lower emissions [1]. As more AVs will be deployed on public roads in the coming decades, their interactions with other road users, such as pedestrians, are expected to increase. One common AV-pedestrian interaction is when pedestrians need to cross a road when the AVs approach the pedestrian. The key to safe and efficient pedestrian-AV interaction is effective communication of vehicles' intentions. With the drivers being inattentive or absent, it is crucial to investigate how AVs can safely interact with pedestrians when pedestrian-driver communication is no longer possible.

Shared space is an urban planning concept that prioritizes people over cars, aiming to create a more open and inclusive public space. In a shared space, traditional road markings, traffic signs, and traffic signals are normally removed to reduce the segregation between different modes of transportation [2]. Interactions among road users rely on social implicit and explicit communications. Shared space calls for redefining the concept of crossing the road, as it involves pedestrians having a path that may come into conflict with the path of an AV, further confounded by the absence of a human driver to interact with and the lack of clear right-of-way rules.

External human-machine interfaces (eHMIs) may be a potential solution to communicate the vehicle's status, intent, and perception to pedestrians in shared space. Investigating AV–pedestrian interaction is complicated, due to the advantages of immersion, high experimental control, and automatic data collection [3], an increasing number of studies use Virtual Reality (VR) to study the effect of eHMI on pedestrian road crossing behavior when interacting with AVs. Numerous VR studies have tested different types of eHMI solutions, including text-based displays [4], [5], graphical displays [6], [7], projections [6], [8], and more [9]. Among different designs of eHMI, research has shown that pedestrians prefer graphical displays on the vehicle and street projects that explicitly indicate the vehicle's intention to yield [10].

While VR provides flexibility to generate diverse traffic scenarios, most existing studies have concentrated on the impact of eHMI on pedestrian crossing behavior in traditional road and crossing scenarios. There is not much exploration of the effect of eHMI in the context of shared-space environments, where there are no overt rights of way. Therefore, while shared space can be an effective approach to improve pedestrian priority in urban areas, it is important to investigate the interaction of pedestrians and AVs and the role of eHMI on pedestrian crossing decisions in a shared-space context to ensure the safety of pedestrians.

The current study investigates the effect of various types of eHMI on pedestrian road crossing behavior with AVs in shared space. During the experiment, pedestrians were asked to cross the road when interacting with an AV under different eHMI conditions, including no eHMI, projected zebra eHMI, and pedestrian eHMI. Both objective (e.g., movement trajectory, gaze point) and subjective (e.g., user experience, trust in AV) data were collected during the VR experiment to study the impact of the different eHMI conditions on the crossing decisions and behavior of pedestrians. In addition, the data collection enables the investigation of the feasibility and effectiveness of a fully immersive and full-motion VR-based approach to study pedestrian-AV interactions.

As one of the pioneering studies that use VR to investigate the impact of eHMI on pedestrian road crossing behavior in shared space, this study contributes to the knowledge gap and provides an initial step toward understanding pedestrian-AV

* Research supported by NWO.

Yan Feng is with the Transport & Planning Department, Delft University of Technology, Delft, 2628 CD, The Netherlands (corresponding author; e-mail: y.feng@tudelftt.nl).

Haneen Farah is with the Transport & Planning Department, Delft University of Technology, Delft, 2628 CD, The Netherlands (e-mail: h.farah@tudelft.nl).

Bart van Arem is with the Transport & Planning Department, he is Senior IEEE Member, Delft University of Technology, Delft, 2628 CD, The Netherlands (e-mail: b.vanarem@tudelft.nl).

interaction in shared space. The findings of this study will provide insights into how AV can be effectively integrated into shared space and understand the role of eHMI in ensuring safe and efficient pedestrian crossings. Moreover, the implementation of free-motion VR allows us to gain a deeper understanding of immersive VR's capability in studying pedestrian-AV interactions.

The remainder of this paper is organized as follows. Section 2 details the VR experiment. Section 3 presents the experimental results and discusses the findings. Section 4 presents the conclusions and future research directions.

## II. METHOD

The current study applied VR experiments to investigate the effect of eHMI on pedestrian road crossing behavior. Ethical approval was obtained from the Human Research Ethics Committee of the Delft University of Technology (Reference ID: 2042). This section presents a detailed description of the experiment method.

### A. Virtual reality environment

The virtual reality environment was developed using Unity Engine based on an existing 3D environment. It represents the Marineterrein area in Amsterdam, Netherlands. The Marineterrein area is an urban district with low-speed motorized movement, narrow street space, and no overt right of way, which is close to the concept of shared space. Besides basic infrastructures (e.g., buildings, roads, and pavements), other environmental elements were also constructed in the virtual environment to resemble realistic urban scenarios, such as trees, grass, and water. An urban soundscape was used to enhance the realism of the virtual environment, which was based on an audio recording of the Marineterrein area. Fig. 1 shows the top view of the virtual environment. The current experiment focused on pedestrian-AV interaction in the T-junction condition (110-meter long and 6-meter wide), indicated by the green line in Fig. 1.

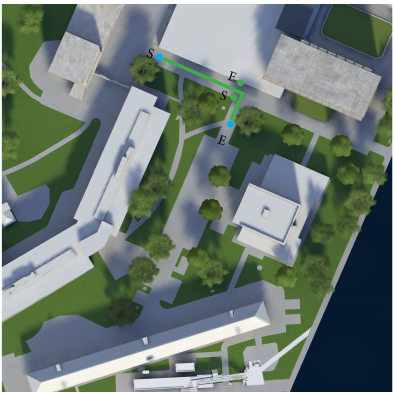

Figure 1. The top view of the virtual environment

### B. Experiment design

To investigate the effect of different types of eHMI on pedestrian road crossing behavior, three levels of eHMI were included, namely none eHMI, eHMI with a pedestrian sign on the AV, and eHMI with projected zebra on the road. The latter two types of eHMI design were chosen to notify pedestrians about the AV's yielding intention. Fig. 2 shows the overview of the tested eHMIs in the current study.

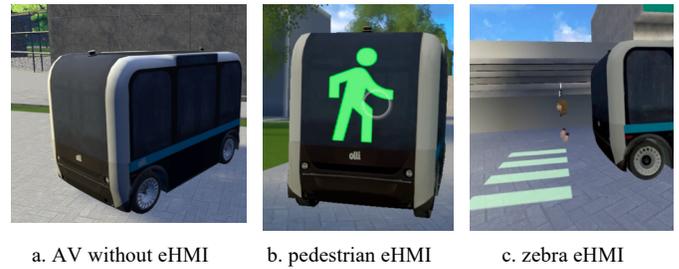

a. AV without eHMI    b. pedestrian eHMI    c. zebra eHMI

Figure 2. The overview of tested eHMIs

For the pedestrian-sign eHMI (Fig. 2b), a green-color pedestrian sign was displayed on the windshield to indicate that the AV intends to yield to the pedestrian and that the pedestrian can cross the road. Another eHMI concept was adapted based on the design of [11] and [6]. When the AV started to yield, it projected a green-color crossing zebra on the road to indicate its intention to yield to the pedestrian (Fig. 2c).

This study used a within-subjects design, in which each participant took part in each of the three experimental conditions exactly once. At the beginning of each scenario, the participant faced the corresponding street. There was a green circle on the ground near the curbside that indicated the starting position of the road-crossing tasks. At the same time, another green circle appeared on the opposite side of the road to indicate the ending position. When the participant stepped into the green circle from the initial position, the AV started to approach the pedestrian from 30 meters away at a speed of 15 km/h in accordance with the speed limit of shared space in the Netherlands. The starting and ending positions of the pedestrian and the AV are indicated by green and blue dots in Fig. 1, respectively. The AV began to decelerate from 15 km/h to 10 km/h with a deceleration rate of 2.5 m/s$^2$ when it was 15 meters from the pedestrian. In both the pedestrian eHMI and zebra eHMI conditions, the AV started to yield and display the eHMI sign at a distance of 5.6 meters from the participant. Once the distance between the pedestrian and AV was equal to or smaller than 1.25 meters, AV came to a complete stop. Participants were instructed to cross the street at the last moment they feel safe to do so. Their task was described as follows: 'Please cross at the last moment you feel safe to cross'. The instruction design was adopted from the study of [12].

### C. Experiment apparatus

During the experiment, two VR headsets were employed, namely an HTC VIVE Pro Eye headset (resolution: 1440 x 1600 pixels per eye, 110 of field-of-view, a 90HZ refresh rate) and an HP Reverb G2 Omnicept headset (resolution: 2160 x 2160 pixels per eye, 114 of field-of-view, a 90HZ refresh rate). Both devices come with built-in eye-tracking sensors and headphones. Fig. 3. shows one participant during the experiment.

To create a realistic and intuitive road crossing experience in shared space, the real-walking locomotion style was adopted. Real-walking locomotion utilizes physical movement to enable movement in the virtual environment under a 1:1 scheme. It means that users can have continuous motion regarding movements and rotations in the real world, which can be matched to the virtual environment. Studies showed that real-walking locomotion generates higher presence sensations, procedure fewer errors in the required task, and is more natural to learn and walk compare to other locomotion styles [13], [14].

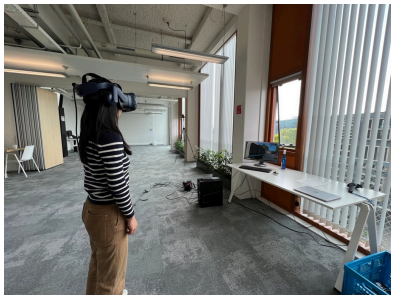

Figure 3. One participant during the VR experiment

### D. Experiment procedure

The experiment procedure includes the following four parts:

1. Introduction: When participants arrived at the experiment location, they were provided written information about the experiments on paper, including the experiment procedure, the explanations of what is AV, the demonstration, and meaning of the tested eHMIs. Following the introduction, participants also read and signed the consent form.

2. Familiarization: Participants were invited to wear the headset and they were immersed in a simple virtual environment to walk around. This part was designed to familiarize participants with navigation in the virtual environment.

3. Official experiment: Once participants felt comfortable and confident enough, the official experiment part started. They were randomly assigned to one experiment scenario and asked to perform the experiment tasks.

4. Filling in the post-questionnaire: After completing the VR experiment, participants were asked to fill in a post-questionnaire. Afterward, participants received around € 20 vouchers as compensation. The entire experiment took less than 1 hour to complete.

### E. Data collection

Within Unity, the participant's movement in the virtual environment was recorded. The data recording started when participants arrived at the first green circle and ended when participants reached the second green circle. The collected data included (1) participant's position (i.e., coordinate x, y, z), (2) head rotation (i.e., roll, yaw, pitch), and (3) gaze point (i.e., coordinate x, y, z). All data were recorded at a frequency of 20 HZ.

The questionnaire included seven parts, namely participant's information, the face validity questionnaire, the Simulator Sickness Questionnaire, the Presence Questionnaire, the Trust in AVs questionnaire, the Perceived behavioral control and risk questionnaire, and the System Usability Scale questionnaire.

### F. Participant's characteristics

In total, 54 participants (27 male and 27 female) aged between 17 and 76 years old ($M$ = 33.63, $SD$ = 14.08) were recruited and took part in the experiment. None of the participants dropped out of the experiment due to motion sickness. The characteristics of the participants are shown in Table 1.

### G. Data analysis

For the objective measures, we focused on determining differences in pedestrian crossing behavior among three eHMI conditions. Five metrics were considered and derived from the movement trajectory data, including crossing initiation time, time before crossing, time to cross, crossing speed, gait variability, and gazing time towards AV. Shapiro–Wilk tests were first performed to examine whether the data is normally distributed. When the normality assumptions were not met for parametric tests, the Friedman test and post hoc Wilcoxon signed-rank tests were performed for each metric. For the subjective measures, the questionnaire data were analyzed to understand the participant's experience of using the VR system and their general attitude toward AV.

TABLE I. DEMOGRAPHIC INFORMATION OF PARTICIPANTS.

| Descriptive information | Category | Number (percentage) |
|---|---|---|
| Highest education level | High school | 2 ( 3.70%) |
| | Associate degree | 3 ( 5.56%) |
| | Bachelor's degree | 27 (50.00%) |
| | Master's degree | 20 (37.04%) |
| | Doctoral degree | 2 ( 3.70%) |
| Previous experience with VR | Never | 17 (31.48%) |
| | Seldom | 25 (46.30%) |
| | Sometimes | 9 (16.67%) |
| | Often | 1 ( 1.85%) |
| | Very often | 2 ( 3.70%) |
| Familiarity with any computer gaming | Not at all familiar | 10 (18.52%) |
| | A-little familiar | 12 (22.22%) |
| | Moderately familiar | 14 (25.93%) |
| | Quite-a-bit familiar | 6 (11.11%) |
| | Very familiar | 12 (22.22%) |
| Familiarity with the Marineterrein area | Not at all familiar | 4 ( 7.41%) |
| | A-little familiar | 3 ( 5.56%) |
| | Moderately familiar | 4 ( 7.41%) |
| | Quite-a-bit familiar | 12 (22.22%) |
| | Very familiar | 31 (57.41%) |
| Familiarity with the concept of automated shuttles | Not at all familiar | 4 ( 7.41%) |
| | A-little familiar | 9 (16.67%) |
| | Moderately familiar | 16 (29.63%) |
| | Quite-a-bit familiar | 16 (29.63%) |
| | Very familiar | 9 (16.67%) |
| Previous experience with automated shuttles | Never | 41 (75.93%) |
| | Seldom | 11 (20.37%) |
| | Sometimes | 1 ( 1.85%) |
| | Often | 1 ( 1.85%) |
| | Very often | 0 ( 0.00%) |

## III. RESULTS AND DISCUSSION

### A. Objective measures

*Crossing initiation time:* Crossing initiation time (CIT) was defined as the duration from the first moment the pedestrian sees the AV until the pedestrian starts to cross the road. CIT is negative when the participant initiates the crossing before seeing the AV, and positive if they initiated crossing after that.

It is interesting to note that not all participants decided to cross until they noticed the AV, 17 (no eHMI condition), 28 (pedestrian eHMI condition), and 18 (zebra eHMI condition) participants decided to cross the road before they paid attention to the AV. Although Friedman test did not find a significant difference in CIT among three eHMI conditions ($\chi^2(2) = 5.11$, $p = 0.078$), Fig. 4 shows the CIT varies across three eHMI conditions. Participants had the lowest CIT for the pedestrian eHMI condition ($M = -0.55$, $SD = 1.94$) and the highest CIT for the zebra eHMI condition ($M = -0.02$, $SD = 2.01$), followed by the no eHMI condition ($M = -0.13$, $SD = 1.91$). The results indicate that participants were more likely to make the crossing decision earlier in the pedestrian eHMI condition.

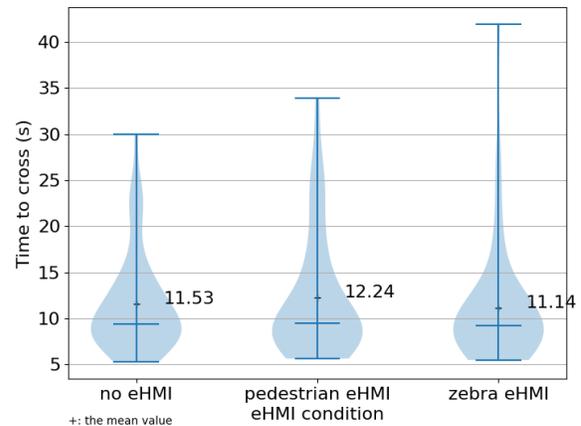

Figure 5.   Violin plot of TTC among different eHMIs.

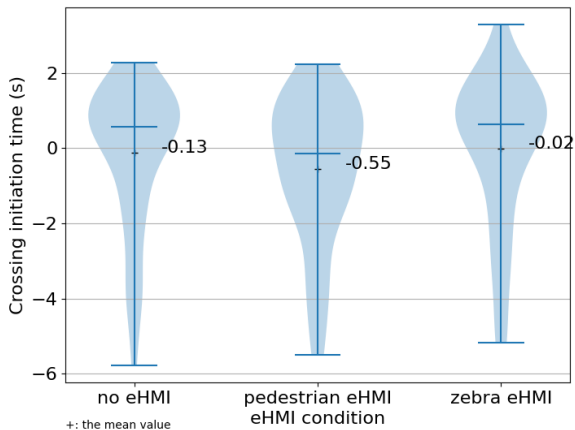

Figure 4.   Violin plot of CIT among different eHMIs.

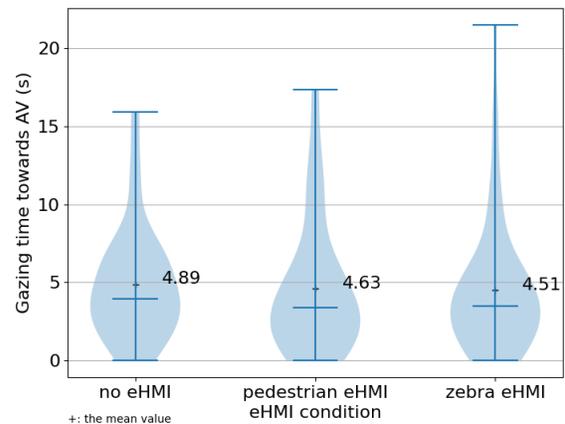

Figure 6.   Violin plot of gazing time towards AV among different eHMIs.

*Time before crossing:* Time before crossing (TBC) is calculated from the start of the experiment until the participant decides to cross. Despite no significant difference in TBC was found among three eHMI conditions ($\chi^2(2) = 4.54$, $p = 0.103$), participants spent the longest TBC in zebra eHMI condition ($M = 1.49$, $SD = 0.70$), followed by no eHMI condition ($M = 1.32$, $SD = 0.65$), and pedestrian eHMI condition ($M = 1.26$, $SD = 0.45$).

*Time to cross:* Time to cross (TTC) is defined as the duration it takes for the participant to reach the opposite of the road from the moment they initiate the crossing. Friedman test finds no significant difference in TTC among three eHMI conditions ($\chi^2(2) = 1.93$, $p = 0.382$). However, the zebra eHMI condition and pedestrian eHMI condition exhibit longer and heavier upper tails in terms of TTC compared to the no eHMI condition, as shown in Fig. 5. This suggests that more participants had a longer TTC and an increased variation in TTC in the zebra and pedestrian eHMI condition.

*Crossing speed:* Crossing speed (CS) is calculated by dividing the total crossing distance by the total time to cross (TTC). No significant differences in CS are found based on Friedman test ($\chi^2(2) = 3.37$, $p = 0.19$). Participants traveled the fastest in the zebra eHMI condition ($M = 0.50$, $SD = 0.14$) than the pedestrian eHMI ($M = 0.48$, $SD = 0.15$) and no eHMI condition ($M = 0.47$, $SD = 0.14$). The results indicate the tendency of participants to have higher crossing speeds in the presence of eHMIs.

*Gazing time towards AV:* Gazing time towards AV is defined as the duration during which participants actively observe the AV. Friedman test showed a significant difference in gazing time among three eHMI conditions: $\chi^2(2) = 8.86$, $p = 0.012$. Wilcoxon signed-rank tests found significant differences between no eHMI and zebra eHMI conditions ($Z = 478.50$, $p = 0.036$), as well as no eHMI and pedestrian eHMI conditions ($Z = 492.50$, $p = 0.048$). The distribution of gazing time of each eHMI condition is presented in Fig. 6. It indicates that the presence of eHMI led to lower gazing time of participants towards the AV.

### B.  Subjective measures

*Realism:* The realism of the VR experience was rated based on a scale that ranges from 1 (not at all realistic) to 5 (completely realistic). The average score of the face validity questionnaire is 3.68 ($SD = 0.57$). Similar scores were found in prior studies that employed VR to study pedestrian road crossing behavior [15]. During the initial encounter with the AV, we even observed one participant feeling uncertain about the AV's action and decided to run to the opposite side of the road. Overall, the rating from participants and observed responses confirmed a moderate realism of the current VR setup.

*Simulation sickness:* The well-established Simulator Sickness Questionnaire [16] was adopted in this study to measure how much simulation sickness participants experienced during the VR experiment. The average score of the total SSQ is 28.40 ($SD$ = 23.23), and no participants reported any discomfort or notable symptoms during the experiment. Based on the rating of SSQ and participant's response, only none and slight symptoms were found in the current study.

*Feeling of presence:* The current study adopted the Presence Questionnaire (PQ) [17] to measure participants' sense of presence. Participants rated 29 items using a 7-point Likert scale. The average total score of PQ in this study is 134.96 ($SD$ = 19.25). This score is slightly higher than [18], [19], which indicates that the participants had a strong sense of presence in the current study.

*Usability:* The System Usability Scale (SUS) questionnaire was adopted to measure the usability of the VR system [20]. It contains 10 items that participants rated from strongly disagree (1) to strongly agree (5). In this study, the average score of SUS is 72.04 ($SD$ = 13.30), which suggests 'good' usability based on the interpretation by [21].

*Trust in AV:* The Trust in AVs questionnaire [7] was used to measure the level of trust in AV. It included 7 items including questions such as 'During the experiment, I trust the automated vehicle to keep its distance from me.' and 'During the experiment, I trust the automated vehicle to drive safely.' Participants rated these questions on a scale from 1 (low trust) and 7 (high trust). The mean score was 4.42 ($SD$ = 1.09) in the current experiment, which shows that participants had a moderate level of trust in the AV.

*Perceived behavioral control and risk:* The Perceived behavioral control (PBC) questionnaire included 2 items, namely 'For me, crossing the road in this way would be …', and 'I believe that I have the ability to cross the road in this way as described in this situation'. Participants rated the questions on a 7-point Likert scale. For the Perceived Risk (PR) questionnaire, participants rated the statement 'Crossing the road in the way as described in this situation would be…' on a scale from 1 (very unsafe) to 7 (very safe). In the current experiment, the mean score of PBC and PR is 5.63 ($SD$ = 0.96) and 5.09 ($SD$ = 1.15), respectively. These are relatively high scores.

*Understanding of eHMI:* Regarding participants' experience with eHMI in the virtual environment, they were asked: 'Based on your experience during the experiment, please choose the most useful and understandable eHMI (external human-machine interface)'. The results show that 32 participants chose the zebra eHMI and 21 participants chose the pedestrian eHMI. This result indicates that zebra eHMI provided participants with a more positive experience and is preferable over pedestrian eHMI.

*C. Discussion*

Regarding the objective measures of pedestrian crossing behavior, the current study revealed that the presence of eHMI influenced gazing time towards AV. This finding indicates that eHMI has a significant impact on directing pedestrians' visual focus toward the AV and consequently their gazing behavior. Additionally, this study found that eHMI did not have a significant impact on crossing initiation time, time before crossing, time to cross, and crossing speed. While this finding aligns with [22], which suggests that an eHMI is not the primary factor influencing pedestrians' crossing behavior, it contradicts studies that demonstrate a positive impact of eHMI on pedestrians' crossing decisions [4], [23]. There could be two reasons for this. Firstly, the studies mentioned above mostly examined the AV operating at a higher speed in mixed traffic, whereas our study focused on an AV operating at a relatively lower speed in shared space. Existing research shows that vehicles' implicit information (e.g., motion cues, deceleration) is the primary factor influencing pedestrian crossing decisions [24], [25]. Therefore, in our study, the low operating speed of the AV takes precedence as a more influential factor compared to the presence of eHMI. Secondly, in the current study, the AV approached pedestrians from a distance of 30 meters. Consequently, participants in our study did not face critical encounters that required immediate crossing decisions. The non-critical nature of the crossing decision was also evident in the crossing initiation time, as many participants chose to begin crossing before directing their attention to the AV. In conjunction with findings from previous studies, this finding implies that eHMI is more effective in scenarios where pedestrians interact with AVs during critical encounters. This finding is consistent with [26], which suggests that eHMI is ineffective when the distance between pedestrians and AV is large.

Regarding the subject measures, this study found that participants experienced the virtual environment with a high level of realism, presence feeling, usability, and a low level of simulation sickness. Meanwhile, participants perceived the AV with a high level of trust, and they crossed the road with a high level of feeling of safety. The positive user experience indicated the effectiveness of the adopted immersive and free-motion VR method to study pedestrian-AV interaction. Moreover, we found the projected zebra eHMI was easier to understand and preferred by participants compared to the pedestrian-sign eHMI. This finding is in line with studies that show eHMI with familiar concepts is preferred by pedestrians [6].

## IV. CONCLUSION

This study investigated pedestrian crossing behavior when interacting with AV with different eHMI designs in shared space using a VR experiment. The key conclusions of this study are twofold. Firstly, this study shows that eHMI might be less efficient in influencing pedestrian crossing decisions during non-critical encounters in shared space when interacting with low-speed AVs, despite the significant impact of eHMI on pedestrian gazing behavior. Secondly, this study shows the feasibility and potential of using immersive and free-motion VR to investigate pedestrian-AV interaction.

This paper presents an initial endeavor to understand pedestrian-AV interaction in shared space, in particular the role of eHMI. There are several limitations of the current study and should be addressed in future research. Firstly, it is crucial to explore pedestrian-AV interaction in shared space with a focus on scenarios involving more critical encounters. It will enable us a better understanding of the dynamics and challenges associated with pedestrian-AV interactions in critical situations. Secondly, it is essential to investigate more realistic and complex interactions between pedestrians and AVs in shared space, such as multiple AVs, pedestrians, various weather and time conditions, as well as other modes of transportation, which could profoundly impact the perception and interpretation of eHMI. Thirdly, future studies should explore the potential correlations among different variables and behavioral metrics presented in the current paper in order to draw more comprehensive conclusions.

ACKNOWLEDGMENT

This work was supported by the project "Safe Interaction of pedestrians and cyclists with automated transport (SIPCAT)" founded by the Netherlands Organization for Scientific Research (NWO). The authors thank The New Base for assisting in developing the VR scenarios. Moreover, we would like to thank AMS Institute and Marineterrein Amsterdam for their facilitation of the SIPCAT project.